\documentclass[
    aps,
	prd,
	reprint,
	amssymb,amsmath,
	nofootinbib,
	superscriptaddress,
	notitlepage
]{revtex4-1} 

\usepackage{graphicx}
\usepackage{dcolumn}
\usepackage{bm}
\usepackage{color}
\usepackage{xspace}
\usepackage{ulem}
\newcommand{\fermi}{\emph{Fermi}\xspace}
\newcommand{\planck}{\emph{Planck}\xspace}
\newcommand{\lat}{\emph{Fermi}-LAT\xspace}
\newcommand{\gr}{$\gamma$-ray\xspace}
\newcommand{\grs}{$\gamma$ rays\xspace}
\newcommand{\msol}{\ensuremath{M_{\odot}}}
\renewcommand{\deg}{\ensuremath{^{\circ}}\xspace}
\newcommand{\ionnew}[2]{\rm{#1\,\textsc{\lowercase{#2}}}}
\newcommand{\hi}{\ionnew{H}{I}\xspace}
\newcommand{\hii}{\ionnew{H}{II}\xspace}

\newcommand\aap{Astron. Astrophys.}
\newcommand\apjs{Astrophys. J. Suppl.}
\newcommand\araa{Ann. Rev. Astron. Astrophys.}
\newcommand\mnras{Mon. Not. R. Astron. Soc.}
\newcommand\jcap{J. Cosmol. Astropart. Phys.}
\newcommand\ssr{Space Sci. Rev.}
\newcommand\pasj{Publ. Astron. Soc. Japan}

\begin{document}
\title{Detection of the Characteristic Pion-Decay Signature in the Molecular Clouds}

\author{Zhao-Qiang~Shen}
\affiliation{Key Laboratory of Dark Matter and Space Astronomy, Purple Mountain Observatory, Chinese Academy of Sciences, Nanjing 210008, China}
\affiliation{University of Chinese Academy of Sciences, Beijing 100012, China}

\author{Yun-Feng~Liang}
\email[]{liangyf@pmo.ac.cn}
\affiliation{Key Laboratory of Dark Matter and Space Astronomy, Purple Mountain Observatory, Chinese Academy of Sciences, Nanjing 210008, China}

\author{Kai-Kai~Duan}
\affiliation{Key Laboratory of Dark Matter and Space Astronomy, Purple Mountain Observatory, Chinese Academy of Sciences, Nanjing 210008, China}
\affiliation{University of Chinese Academy of Sciences, Beijing 100012, China}

\author{Xiang~Li}
\email[]{xiangli@pmo.ac.cn}
\affiliation{Key Laboratory of Dark Matter and Space Astronomy, Purple Mountain Observatory, Chinese Academy of Sciences, Nanjing 210008, China}

\author{Qiang~Yuan}
\affiliation{Key Laboratory of Dark Matter and Space Astronomy, Purple Mountain Observatory, Chinese Academy of Sciences, Nanjing 210008, China}
\affiliation{School of Astronomy and Space Science, University of Science and Technology of China, Hefei, Anhui 230026, China}

\author{Yi-Zhong~Fan}
\email[]{yzfan@pmo.ac.cn}
\affiliation{Key Laboratory of Dark Matter and Space Astronomy, Purple Mountain Observatory, Chinese Academy of Sciences, Nanjing 210008, China}
\affiliation{School of Astronomy and Space Science, University of Science and Technology of China, Hefei, Anhui 230026, China}

\author{Da-Ming~Wei}
\affiliation{Key Laboratory of Dark Matter and Space Astronomy, Purple Mountain Observatory, Chinese Academy of Sciences, Nanjing 210008, China}
\affiliation{School of Astronomy and Space Science, University of Science and Technology of China, Hefei, Anhui 230026, China}

\date{\today}

\begin{abstract}

The \gr emission from molecular clouds is widely believed to have a hadronic origin, while unequivocal evidence is still lacking.
In this work, we analyze the \lat Pass 8 publicly available data accumulated from 2008 August 4 to 2017 June 29 and report the significant detection of a characteristic $\pi^0$-decay feature in the \gr spectra of the molecular clouds Orion~A and Orion~B.
This detection provides a direct evidence for the hadronic origin of their \gr emission.

\end{abstract}

\maketitle

\section{\label{sec:intro}Introduction}
Molecular clouds are one of the few GeV emission sources known in 1980s \citep{Caraveo1980}.
There is a tight linear correlation between the \gr emission and total gas distribution \citep{Bloemen1984,Digel1995,Abdo2009,Ackermann2012a}.
The \gr emission from molecular clouds is widely believed to be mainly from the hadronic process.
Hence the detected \gr emission from some nearby molecular clouds can be used to deduce the cosmic ray (CR) spectra.
Different from most direct measurements by detectors near Earth (the only exception is Voyager 1 that has crossed the heliopause into the nearby interstellar space on 2012 August 25 \citep{Stone2013,Cummings2016}, which has measured the local interstellar proton spectrum below $\sim 300~{\rm MeV}$), such indirect measurements do not suffer from the solar modulation and may provide the local interstellar spectra of the CRs \citep{Aharonian2001, Casanova2010}.
Thanks to the successful performance of \lat \citep{FermiLAT}, the \gr emission properties of interstellar gas have been widely examined and our understanding of the ``local" CR distribution has been revolutionized \citep{Abdo2009,Ackermann2011a,Neronov2012,Ackermann2012a, Ackermann2012b,Ackermann2012c,Yang2014,Tibaldo2015,Casandjian2015,Yang2016,FermiGDE,Neronov2017}.
Though these progresses are remarkable, the characteristic $\pi^0$-decay signature in the molecular clouds has not been reported yet and consequently the direct evidence for the hadronic origin is still lacking.
In this work, we aim to detect such a signature in two ``nearby'' giant molecular clouds (Orion~A and Orion~B).

Orion~A and Orion~B have masses of $\gtrsim 8\times10^4\msol$ and locate at $\sim 450~{\rm pc}$ from the Earth \citep{Wilson2005}.
Their great mass, proximity as well as the favorable position in the sky (high latitude, away from the Galactic center and the Fermi Bubble \citep{Su2010}), render them ideal targets for \gr observation.
High energy emission from the direction of Orion cloud complex was first detected by the COS-B satellite \citep{Caraveo1980, Bloemen1984}, and further by EGRET \citep{Digel1995, Digel1999} and \lat \citep{Ackermann2012a}.
The shapes of the \gr spectra associated with Orion~A and Orion~B are similar to those in the Gould Belt, implying that these two clouds are ``passive'' CR detectors \citep{Neronov2012}.

Taking advantage of the increased acceptance and improved angular resolution of Pass 8 data \citep{LatPass8} and the latest multi-wavelength observation of the interstellar medium (ISM) tracers, the \gr analysis of Orion molecular clouds is revisited in this work.
We will derive the \gr emissivities of Orion~A and Orion~B from 60~MeV to 100~GeV, discuss the systematic uncertainties, examine the structures in the spectra, and determine whether the $\pi^0$ decay mainly accounts for the \gr emission of the molecular clouds.
Special attention is paid to the spectra below 100~MeV which are not covered in the previous \lat analyses on individual molecular clouds \citep{Neronov2012, Ackermann2012a, Yang2014, Neronov2017} and are important to pin down the nature of the emission.

\section{\label{sec:analysis}Data analysis}

\subsection{\label{sec:analysis::gr_data}\gr data}
The Large Area Telescope (LAT) onboard the \emph{Fermi} satellite is a pair-conversion \gr telescope, covering a wide energy range from 20~MeV to more than 300~GeV \citep{FermiLAT,LatOnOrbit}.
We take the \lat Pass~8 CLEAN data set (P8R2\_CLEAN\_V6) to supress the contamination from the residual cosmic ray background \citep{LatOnOrbit,LatPass8}.
Observations accumulated from 2008 August 4 to 2017 June 29 (\fermi mission elapsed time (MET) from 239557417 to 520388318) with energies from 60~MeV to 100~GeV are selected.\footnote{\url{ftp://legacy.gsfc.nasa.gov/fermi/data/lat/weekly/photon/}}
Because of the poor angular resolution for the events below $\sim 2$ GeV and small statistics at higher energies, we construct following two data sets.
For the low energy data set, the front-converting events from 60~MeV to 2~GeV are chosen, while those with the zenith angle larger than 90\deg are excluded to reduce emission from the Earth's albedo.
For the high energy data set, both the front- and back-converting events between 1~GeV and 100~GeV with the zenith angle less than 95\deg are picked.
An overlap from 1~GeV to 2~GeV in the high energy data set is kept merely to check the consistency between the results from the two data sets, and ignored in the Sect.~\ref{sec:discussion}.
Then, the quality-filter cut (DATA\_QUAL$>$0)\&\&(LAT\_CONFIG$==$1), which just keeps the data collected in science mode and removes those associated with the solar flares or particle events, is applied to both data sets.

\begin{figure*}
    \centering
    \includegraphics[width=0.49\textwidth]{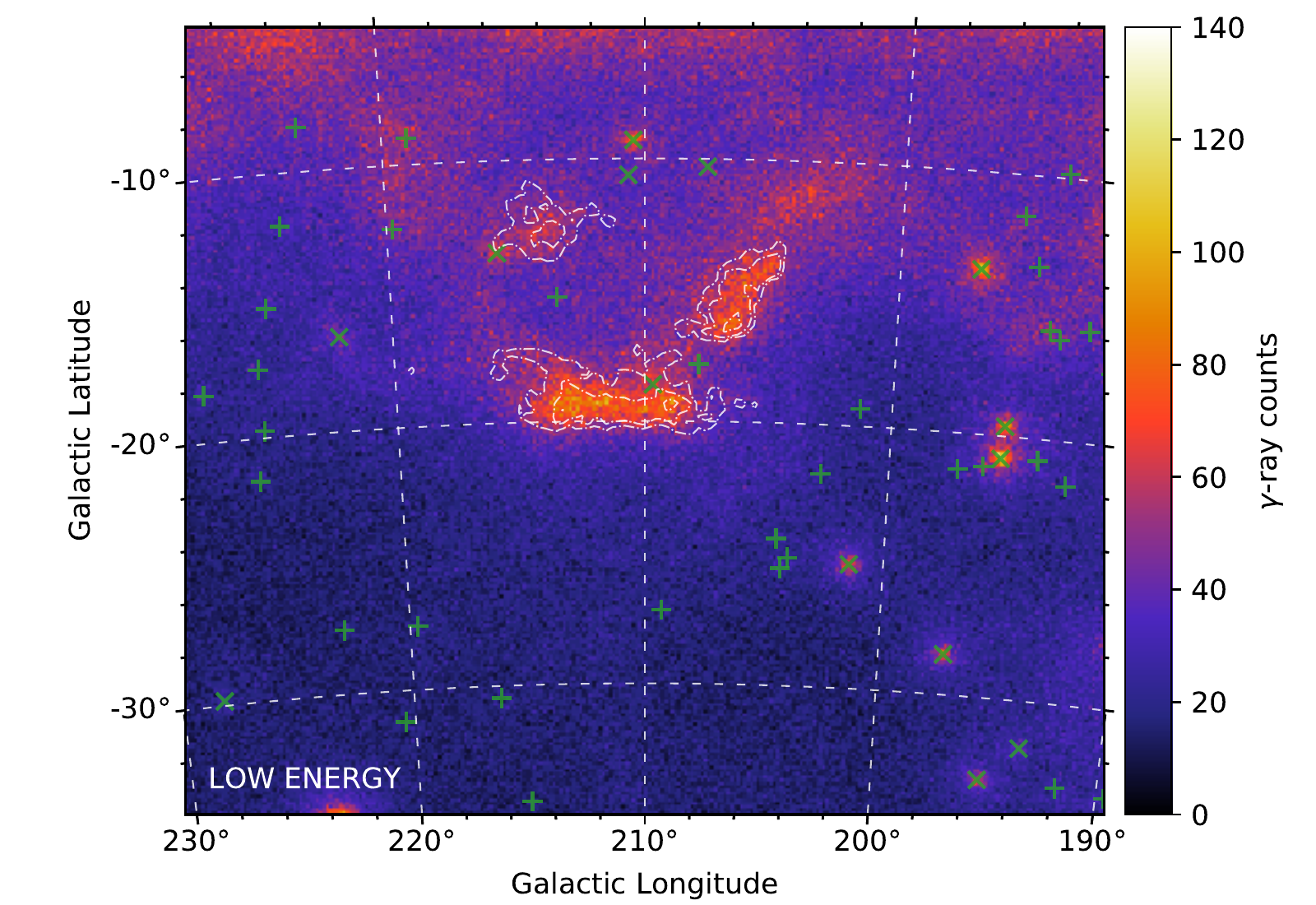}
    \includegraphics[width=0.49\textwidth]{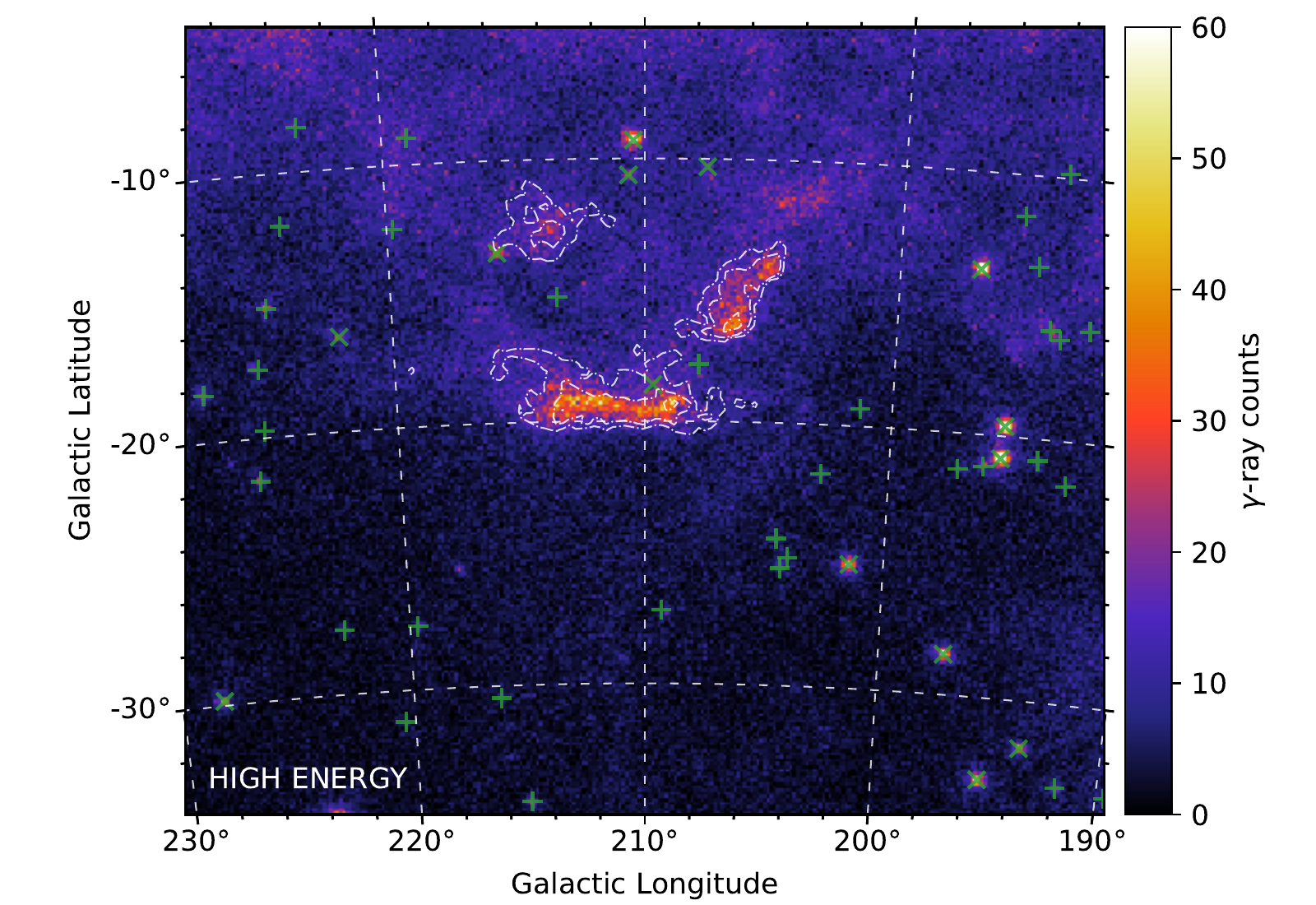}
    \caption{\label{fig:cmap}
        The count maps for \grs in the energy range between 60~MeV and 2~GeV (left panel) and between 1~GeV and 100~GeV (right panel) within the ROI.
        The point sources which are modelled separately in the analysis are drawn as X markers, while others are marked as crosses.
        White dot-dashed contours represent the CO image of molecular clouds Orion~A (lower), Orion~B (upper right) and Mon~R2 (upper left).
    }
\end{figure*}

We define the region of interest (ROI) as a rectangular region in a plate carr\'{e}e projection in Galactic coordinates, which is centered at the middle point of Orion~A ($\ell=210\deg, b=-20\deg$) and consists of $280 \times 240$ pixels with roughly $0.125\deg$ bin size,\footnote{
    To be accurate, $0.125\deg$ width for the pixel in the center of ROI.
} with which 95\% of the \grs originated from the Orion~A in 60~MeV is included in the analysis.
The count maps for two data sets are shown in the Fig.~\ref{fig:cmap}, in which the \gr emission from the giant molecular clouds Orion~A, Orion~B and Mon~R2 are clearly visible.
To build binned count cubes, 12 logarithmically spaced energy bins between 60~MeV and 2~GeV are adopted for the low energy data set, while for the high energy one we split the data into 11 energy bins as shown in Tab.~\ref{tab:emiss}.

The data selection as well as the convolution of the models with the instrument response functions (IRFs) in Sect.~\ref{sec:analysis::gr_analysis} are performed with the latest (v10r0p5) Fermi Science Tools.\footnote{\url{http://fermi.gsfc.nasa.gov/ssc/data/analysis/software/}}

\subsection{\label{sec:analysis::gr_component}\gr radiation components}
The \gr sky observed by \lat can be modelled with a combination of Galactic diffuse components, one isotropic component and a number of point-like or extended sources \citep{3FGL,FermiGDE}.
We will introduce the baseline templates used in Sect.~\ref{sec:analysis::gr_analysis} in this subsection, and leave the systematic uncertainties to Sect.~\ref{sec:analysis::systematic}.
In order to include at least 68\% photons at the edge of the ROI, we define a source region, i.e. the region within which the sources are accounted for, with the Galactic longitude from $182.5\deg$ to $237.5\deg$ and latitude between $-45\deg$ and $5\deg$.

\grs generated by the interaction of energetic CRs with interstellar gas contribute to the Galactic diffuse emission.
High-energy photons will be produced through the decay of $\pi^0$ mesons resulting from hadron collisions, or through the electron bremsstrahlung.
The \gr intensity from these processes is proportional to the column density of gas, if the CRs distribute uniformly in the interstellar gas.
Various observations have revealed the uniformity of CR in atomic gas and molecular clouds above 100~MeV (e.g. \cite{Abdo2009,Abdo2010a,Ackermann2011a}), so the spatial templates derived from the column density of different phases of gas can be made as a starting point of \gr analysis.
Moreover, we assume that helium and heavier elements in the gas are uniformly mixed with hydrogen \citep{Casandjian2015}.

\begin{figure*}
    \centering
    \includegraphics[width=0.49\textwidth]{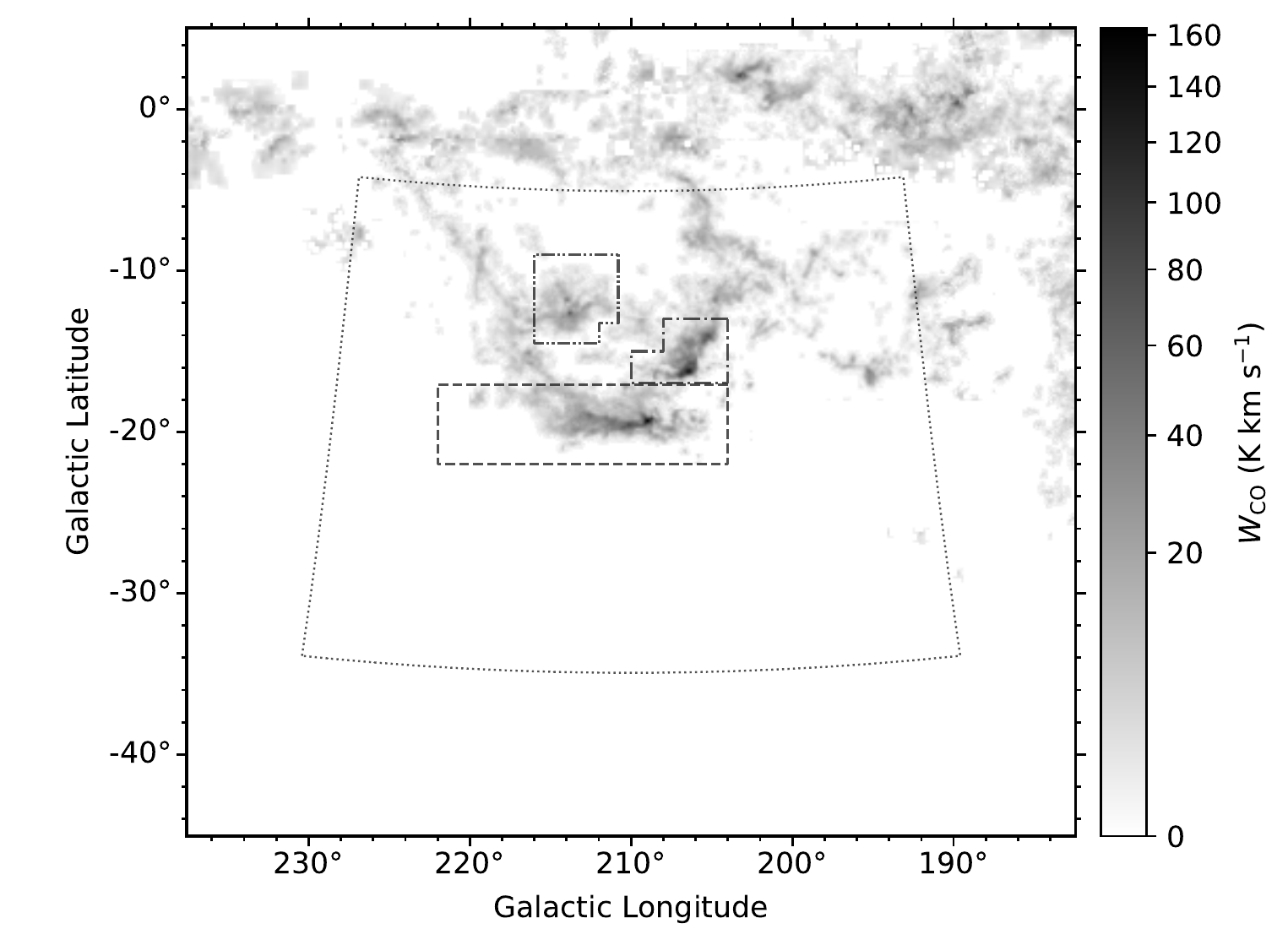}
    \includegraphics[width=0.49\textwidth]{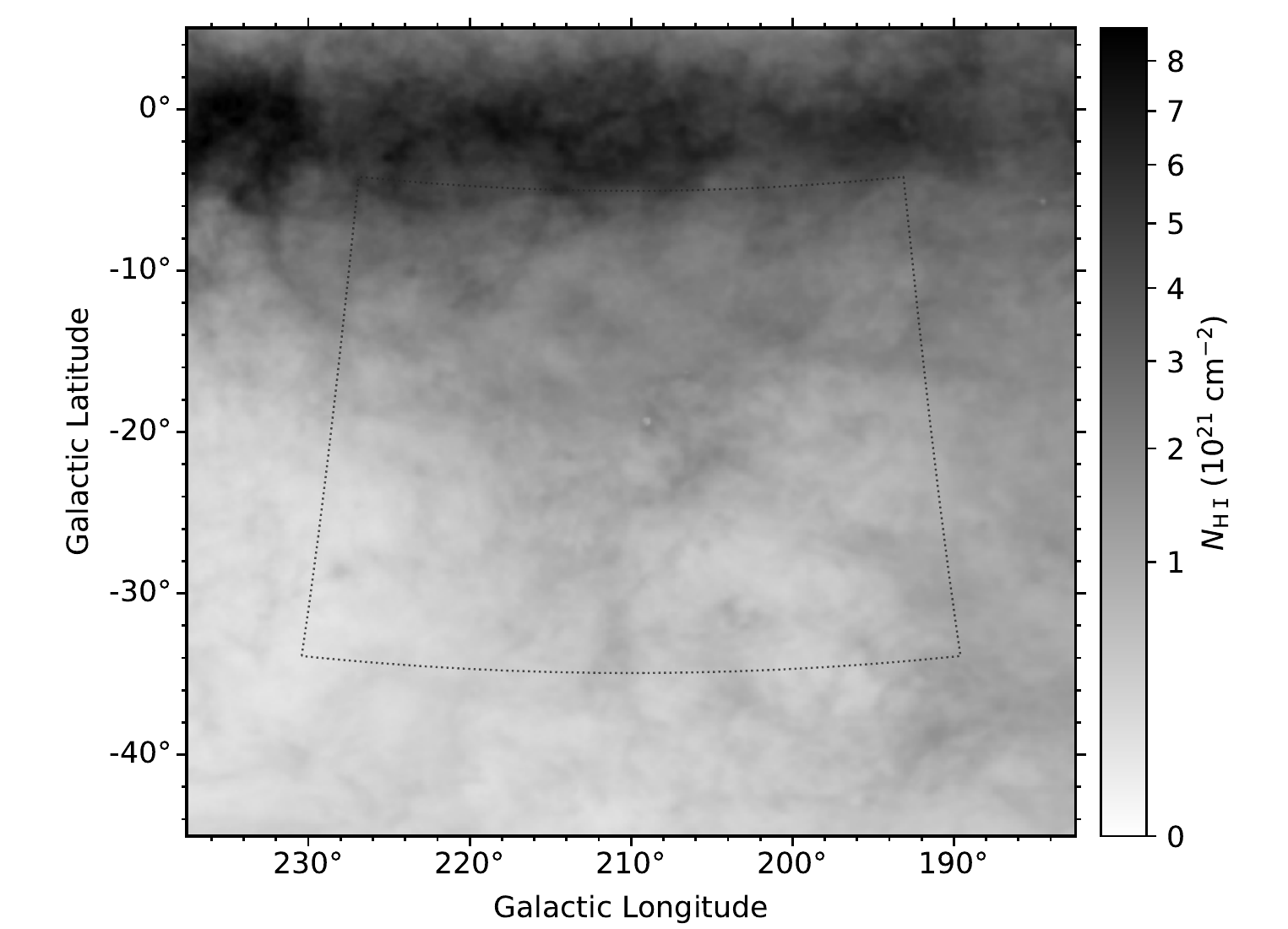}
    \includegraphics[width=0.49\textwidth]{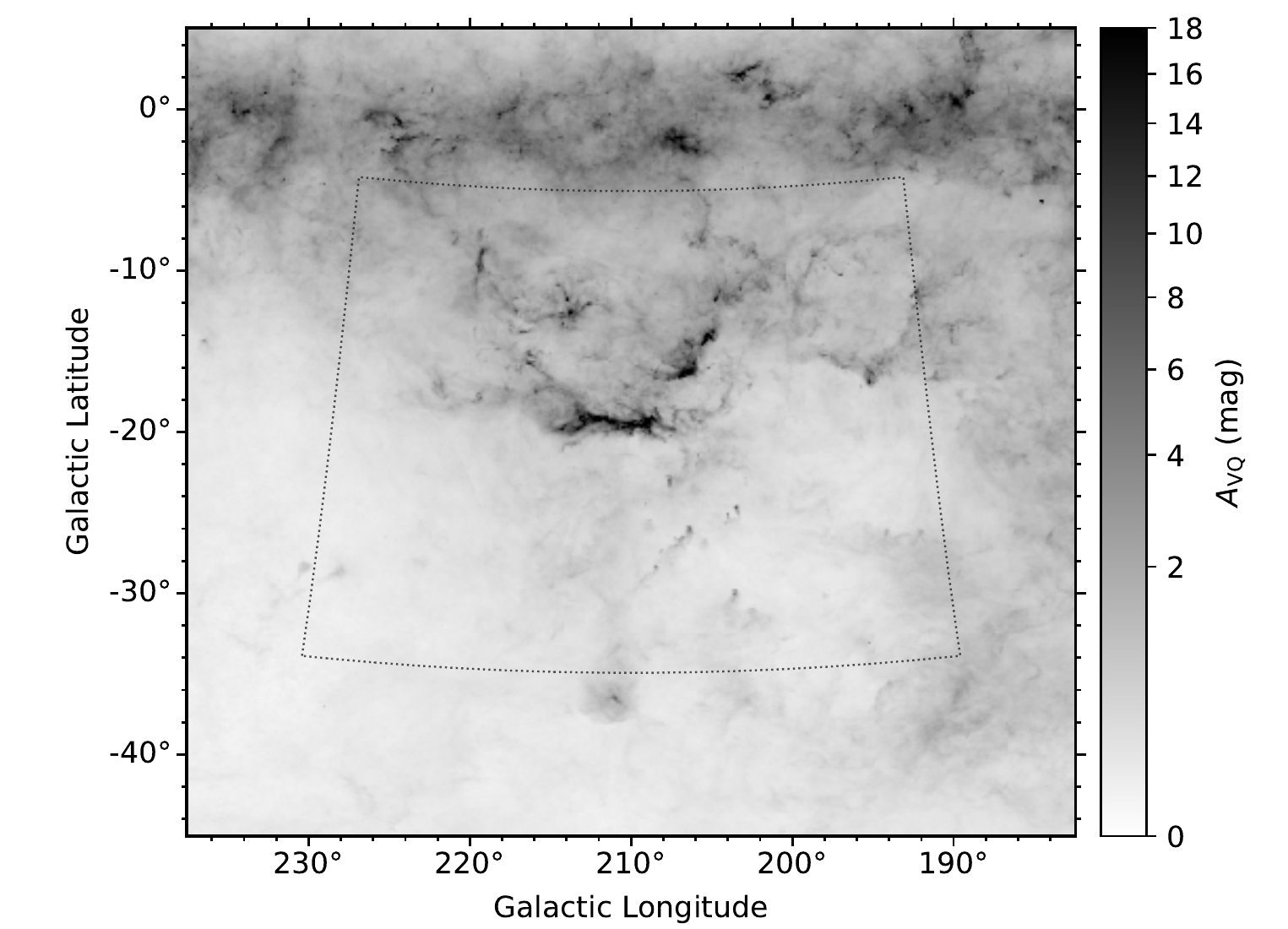}
    \includegraphics[width=0.49\textwidth]{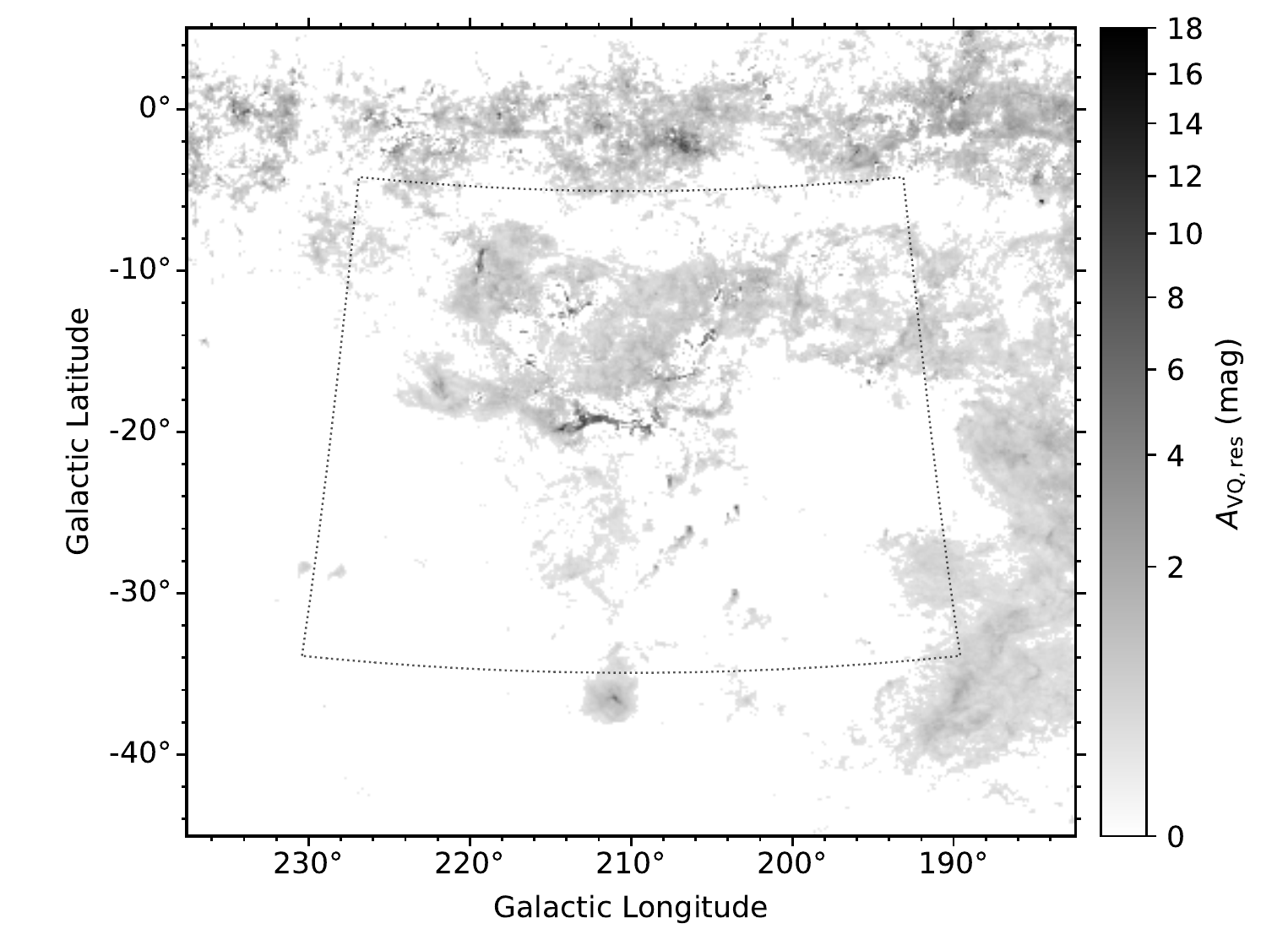}
    \caption{\label{fig:gr_gas}
        The spatial templates of different phases of gas.
        The dotted line illustrates the ROI of the \gr analysis.
        The top left panel is the $W_{\rm CO}$ map, in which the dashed, dash-dotted and dash-double-dotted lines indicate the regions corresponding to Orion~A, Orion~B and Mon~R2, respectively.
        The top right panel is the template for the column density of atomic hydrogen under the optically thin assumption.
        The lower left panel is the renormalized $A_{\rm VQ}$ extinction map, while the lower right panel represents the derived $D_{\rm DNM}$ map.
    }
\end{figure*}

H$_2$ is the most abundant interstellar molecule.
However, due to the lack of a permanent electric dipole, cold hydrogen molecules can not be directly observed.
The 2.6~mm line coming from the rotational transitions $J=1\to0$ of $^{12}$CO is wildly used as the tracer of H$_2$ \citep{Dame2001}.
The CO data observed by the Center for Astrophysics (CfA) 1.2~m telescope \citep{Dame2001,Wilson2005} and further denoised with the moment-masking method \citep{Dame2011} are adopted in this work.
We integrate the brightness temperature over the velocity in each pixel to construct a $W_{\rm CO}$ map of the source region, which is empirically proportional to the H$_2$ column density by a factor $X_{\rm CO}$.
This map is binned with pixel size of 0.125\deg with those sparsely sampled pixels linearly interpolated using the data from the positive ones among nearest 8 pixels.
Since there may be different $X_{\rm CO}$ and CR intensity in different molecular clouds \citep{Grenier2015,FermiGDE,Yang2016}, we split the $W_{\rm CO}$ map into regions of Orion~A ($204\deg<l<222\deg$, $-22\deg<b<-17\deg$), Orion~B ($204\deg<l<210\deg$, $-17\deg<b<-13\deg$, but exclude uncorrelated regions in $208\deg<l<210\deg$, $-15\deg<b<-13\deg$), Mon~R2 ($210.8\deg<l<216.0\deg$, $-14.5\deg<b<-9.0\deg$, but exclude $210.8\deg<l<212.0\deg$, $-14.5\deg<b<-13.3\deg$) \citep{Wilson2005}.
We also separate the rest part of Orion-Monoceros complex ($200\deg<l<226\deg$, $-30\deg<b<-4\deg$) with other high latitude clouds ($b<-5\deg$) and the Galactic Plane (the rest part).
The $W_{\rm CO}$ map as well as some defined regions can be seen in Fig.~\ref{fig:gr_gas}.

Neutral atomic hydrogen, more spatially extended than molecule, is also a vital part of the interstellar gas.
Its spatial distribution can be well mapped at radio wavelengths with 21-cm hyperfine structure line.
Recently \hi 4$\pi$ survey (HI4PI) published the most sensitive all-sky \hi survey result, which provides the brightness temperature of the \hi line with a grid of $0.083\deg$ and covers the local standard of rest velocities between $-600~{\rm km~s^{-1}}$ and $600~{\rm km~s^{-1}}$ \citep{HI4PI}.
Under the approximation of small optical depth (see Sect.~\ref{sec:analysis::systematic} for the uncertainty due to the choice of the optical depth), we calculate the column density map of \hi by integrating spectroscopic data over the velocity using the formula $N_{\hi}(l,b)= C\int {\rm d}v\,T_{\rm B}(l,b,v)$, where $T_{\rm B}(v)$ is the brightness temperature profile and $C=1.823\times10^{18} {\rm atoms~cm^{-2}~(K~km~s^{-1})^{-1}}$ \citep{HI4PI}.
Foreground absorptions around $l=209.0\deg$ and $b=-19.4\deg$ are found in the source region, for which the column densities are replaced with the interpolated ones from the nearby positive pixels.
The processed \hi map is shown in Fig.~\ref{fig:gr_gas}.
We are aware that some extragalactic objects exist in the HI4PI data.
However, since our source region do not overlap those prominent objects such as the Magellanic Clouds, M31 and M33, the column density map still traces Milky Way's \hi quantitatively \citep{HI4PI}.

The ionized gas is also a component of interstellar gas.
Due to the low density of \hii in our source region, it however has been ignored in the following analysis (see also \cite{Ackermann2012a}).

Both the dust and the \gr emission traces the total gas.
By analyzing both tracers, gas invisible in the \hi and CO is unveiled \citep{Grenier2005}.
The dark gas, also known as dark neutral medium (DNM), may be made up of atomic gas which is overlooked due to the approximation of \hi opacity, or molecular gas which is not associated with CO \citep{Grenier2015}.
This component has approximately the same mass as the molecular gas in a sample of $10^4-10^6\msol$ clouds \citep{Grenier2015}, so we need to take it into account.
The structured component other than \hi and CO in the dust map is the DNM template we want.
To extract this component, we first use the $A_{\rm VQ}$ extinction map\footnote{\url{http://pla.esac.esa.int/pla}} as our dust tracer template, which is yielded by fitting the data recorded by \planck, IRAS and WISE to the dust model \citep{Draine2007} and renormalizing to the quasi-stellar objects observed in the Sloan Digital Sky Survey \citep{Planck_Avq}.
Because the properties of dust differ between the diffuse and dense environments \citep{Ade2015,Mizuno2016,Remy2017}, we set the maximum $A_{\rm VQ}$ values in the map to $15.5~{\rm mag}$ which corresponds to $E(B-V)=5~{\rm mag}$ \citep{Ackermann2012b}.
This map is presented in the lower left panel of Fig.~\ref{fig:gr_gas}.
Assuming a uniform dust-to-gas ratio and mass emission coefficient, we can then model the dust tracer map using a linear combination of the gas maps we made above, i.e.
\begin{equation}
    M(l,b)
        = \sum\nolimits_{i=1}^6 y_{{\rm CO},i} \, W_{{\rm CO},i}(l,b)
        + y_{\rm \hi} \, N_{\rm \hi}(l,b)
        + y_{\rm iso},
    \label{eqn:dust_model}
\end{equation}
where the index $i$ labels one given CO map (in total there are six).
The $y_{\rm iso}$ term is introduced to account for the residual noise of dust map in the zero level \citep{Ade2015}.
The goodness of the fit to the dust map\footnote{
    We mask the regions with $A_{\rm VQ}>15.5~{\rm mag}$ in the dust fitting, but leave them in the residual map to model DNM component.
} is governed by
\begin{equation}
    \chi_{\rm dust}^2 = \sum_{l,b} \frac{\left[ D(l,b)-M(l,b) \right] ^2}{\sigma^2(l,b)},
    \label{eqn:chi2_dust}
\end{equation}
where $D(l,b)$ stands for the dust map, which is $A_{\rm VQ}$ here, and $\sigma(l,b)$ is defined to be proportional to $D(l,b)$ \citep{Ade2015,Tibaldo2015,Remy2017}.
To extract the map containing significant positive residuals, we build a histogram of the residual map, fit a Gaussian curve to the core of the distribution, and set those pixels with residuals below the $+5\sigma$ of the Gaussian profile to be zero.
The derived map is the DNM template ($D_{\rm DNM}$ map, or $A_{\rm VQ, res}$ here) shown in Fig.~\ref{fig:gr_gas}.

The inverse Compton (IC) radiation of high energy electrons also contributes to the Galactic diffuse emission.
Since no direct observational template is available for it, theoretical models calculated by the CR propagation code {\tt GALPROP} \citep{Galprop} are often adopted in diffuse \gr analysis.
The same IC component in the standard \lat Galactic interstellar emission model, which has the GALDEF identification ${\rm ^SY ^Z6 ^R30 ^T150 ^C2}$, is adopted in the baseline model \citep{FermiGDE}.
This model is made under the assumptions that the distribution of CR sources is proportional to that of pulsars given by \cite{Yusifov2004} and CRs propagate within the Galactic halo with 6~kpc half height and 30~kpc galactocentric radius in the diffusive reacceleration model \citep{Ackermann2012b}.
We use the {\tt GALPROP}\footnote{\url{https://galprop.stanford.edu/}} code (v54.1.984) to produce the IC files used in the Fermi Science Tools \citep{Strong2000,Porter2008}.

The last diffuse component distributes isotropically in the sky, which comprises of the unresolved extragalactic emissions, residual Galactic foregrounds and CR-induced backgrounds \citep{Ackermann2015}.
Since we use the templates above instead of the standard \lat Galactic interstellar emission model, the spectrum of the isotropic component may also be different from the standard one.
So when performing the bin-by-bin analysis, we let the spectral parameters of the isotropic component vary.

We also consider the sources listed in the \lat third source catalog (3FGL) \citep{3FGL} within the source region.
Since the 3FGL is produced based on the first 4 years of observation, some weak or transient sources not included in the 3FGL may appear in our data.
We ignore these sources due to their relatively small contribution.\footnote{
    The weak point sources ($<15\sigma$) in the 3FGL contribute to only $\lesssim 3\%$ of the photons within the ROI, let alone those not in the catalog.
} Also, there are some sources in the catalog that are considered to be potentially confused with interstellar emission \citep{3FGL}.
To be conservative, we remove the sources in the regions of Orion~A, Orion~B and Mon~R2.
Because the spectral parameters of point sources in the 3FGL are derived with the standard Galactic emission model \citep{3FGL}, using the parameters in the catalog directly may lead to bias.
We model both the bright ($\geq 15\sigma$) sources within the ROI and the sources overlapping the Orion~A, Orion~B or Mon~R2 with the significance $\geq 5\sigma$ (drawn as X markers in Fig.~\ref{fig:cmap}) as individual templates.
And the rest ones are merged into a single template using {\tt gtmodel} to limit the number of free parameters in the fittings.

\subsection{\label{sec:analysis::gr_analysis}\gr analysis procedure and baseline results}
The \gr intensity $I_{\gamma}$ in the direction $(l,b)$ at the energy $E$ can be expressed as
\begin{eqnarray}
    I_{\gamma}(l,b,E)
        &=& \sum\nolimits_{i=1}^6 q_{{\rm CO},i}(E) \, W_{{\rm CO},i}(l,b) \nonumber\\
        &+& q_{\rm \hi}(E) \, N_{\rm \hi}(l,b)
         +  q_{\rm DNM}(E) \, D_{\rm DNM}(l,b) \nonumber\\
        &+& x_{\rm IC}(E) \, I_{\rm IC}(l,b,E)
         +  x_{\rm iso}(E) \, I_{\rm iso}(E) \nonumber\\
        &+& x_{\rm ps}(E) \, \sum\nolimits_k^{n_{\rm ps,nf}} S_k(E) \, \delta(l-l_k,b-b_k) \nonumber\\
        &+& \sum\nolimits_j^{n_{\rm ps,f}} S_j(E;\theta_j) \, \delta(l-l_j,b-b_j),
    \label{eqn:gr_model}
\end{eqnarray}
where $q$ stands for the \gr emissivity of the gas, and $x$ is the scaling factor for the mapcube model.
$I_{\rm iso}(E)$ is the tabulated isotropic spectrum for standard point-source analysis ({\tt iso\_P8R2\_CLEAN\_V6\_FRONT\_v06.txt} and {\tt iso\_P8R2\_CLEAN\_V6\_v06.txt} for low energy and high energy data respectively), and $S(E;\theta)$ is the spectrum for a point source with the spectral parameters $\theta$.
$n_{\rm ps,f}$ represents the number of point sources with spectral parameters freed, and the number of remaining point sources which are merged into a single template is denoted as $n_{\rm ps,nf}$.\footnote{
    ``ps,f'' and ``ps,nf'' represent point sources with spectral parameters freed and not freed, respectively.
}
The above model incorporating the baseline templates and standard \lat IRFs is our baseline model.

After convolving the intensity with the \lat IRFs using {\tt gtsrcmaps}, we perform binned likelihood analysis by minimizing the function $-2\,\ln [\mathcal{L}(q,x,\theta)]$ with the MINUIT algorithm \citep{MINUIT}, where the likelihood $\mathcal{L}$ has the same form as that implemented in {\tt gtlike}.\footnote{\url{https://fermi.gsfc.nasa.gov/ssc/data/analysis/documentation/Cicerone/}}
To avoid overfitting of the model, we first perform a global fit and then a bin-by-bin analysis.
In the global fit, we choose the LogParabola spectral type for all the scaling factors and emissivities except that of the isotropic, the merged weak point sources and the CO component for the Galactic Plane.
For the isotropic one, we use a constant scaling factor and let the normalization change, while for the merged template of weak sources, we choose a power law model and optimize its normalization and index.
For the Galactic CO component, we adopt the spectral shape from \cite{Casandjian2015} and free its normalization only since most of it is outside the ROI.
Concerning the individual point sources, the prefactors and spectral indices of those with significance $\geq 20\sigma$ within the ROI and those $\geq 15 \sigma$ sources overlapped with Orion~A, Orion~B or Mon~R2 are optimized.
For other ones, we only fit their normalizations.
After the global fit, we freeze all the spectral parameters of the Galactic Plane component as well as those point sources with test statistic (TS) value $<400$,\footnote{
    The TS is defined to be $-2\,{\rm ln}(\mathcal{L}_{\rm max,0}/\mathcal{L}_{\rm max,1})$ \citep{Mattox1996}, in which $\mathcal{L}_{\rm max,1}$ and $\mathcal{L}_{\rm max,0}$ are the best-fit likelihood values of the alternative hypothesis and null hypothesis, respectively.
} and only fit the normalizations of other components in a bin-by-bin way.
It should be noticed that no energy dispersion corrections have been made in our fits.

\begin{figure*}
    \centering
    \includegraphics[width=0.49\textwidth]{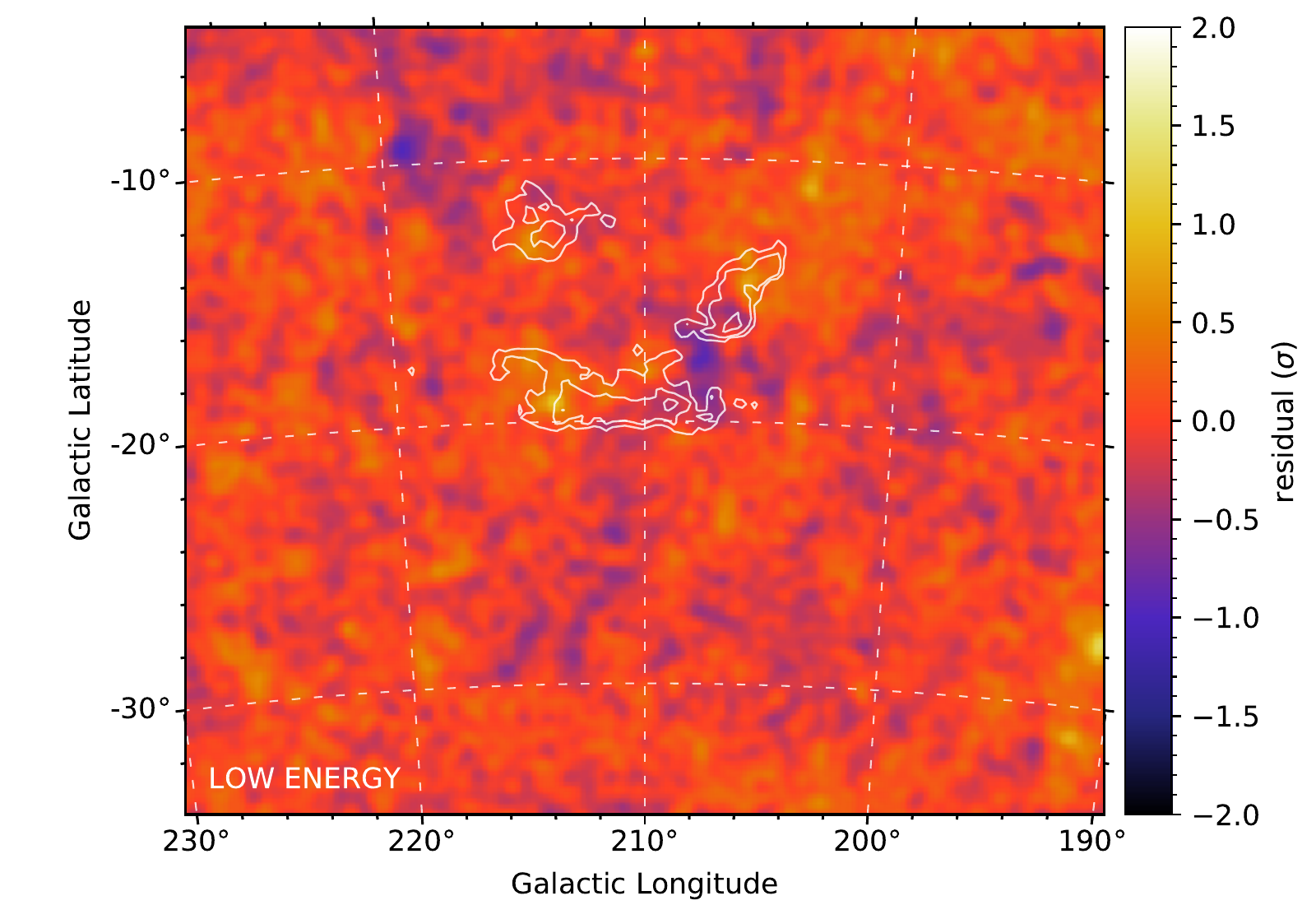}
    \includegraphics[width=0.49\textwidth]{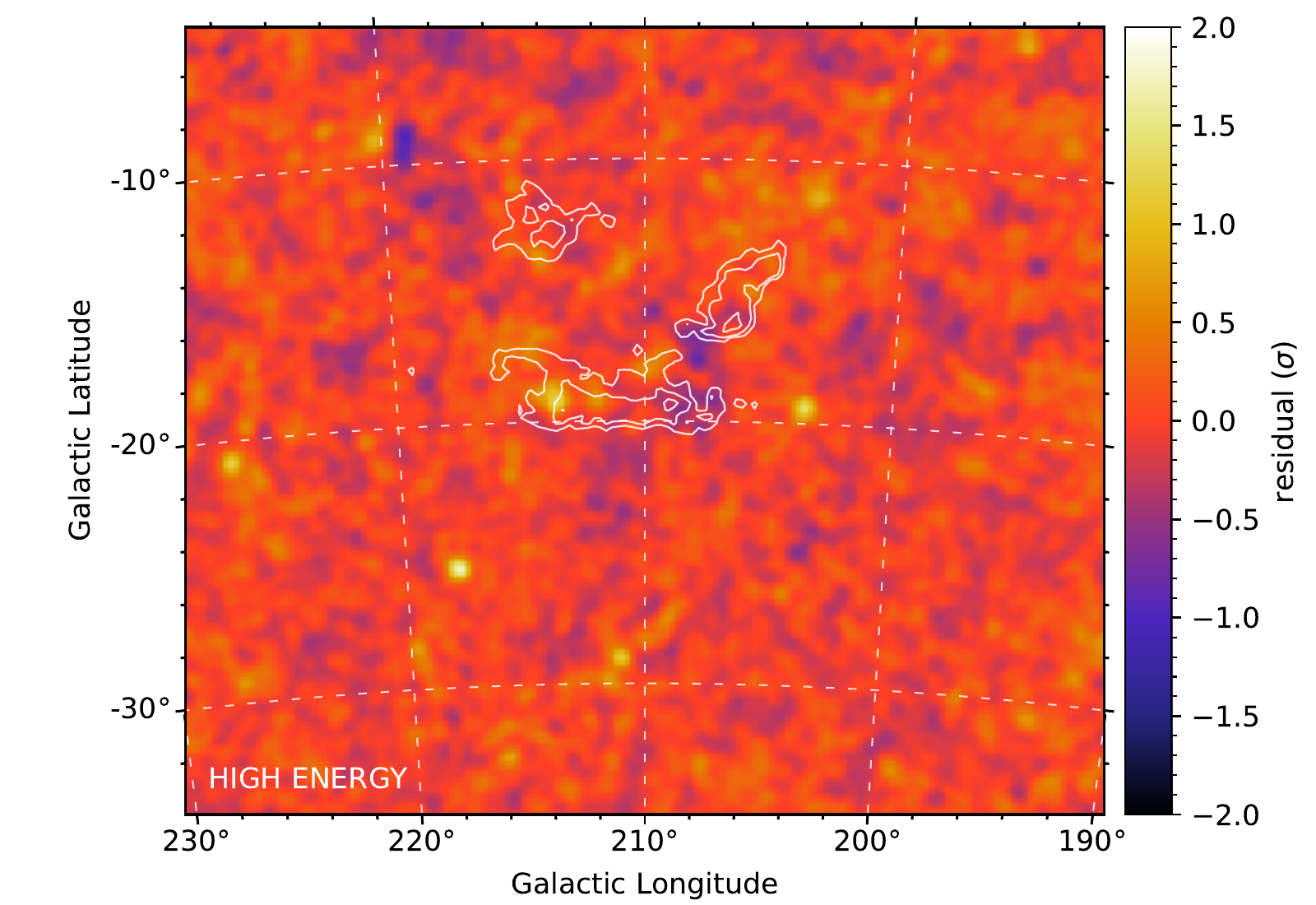}
    \caption{\label{fig:residual}
        The residual maps in the energy range between 60~MeV and 2~GeV (left panel) and between 1~GeV and 100~GeV (right panel) within the ROI.
        The residuals are in the unit of standard deviations, which are calculated by dividing the differences between the observed counts and the predicted counts to the square root of the predicted counts.
        To decrease statistical fluctuation in the residual maps, we smooth both the observed and predicted count maps using a Gaussian kernel of $\sigma=0.25\deg$.
        The white contours indicate the CO images of Orion~A, Orion~B, Mon~R2.
    }
\end{figure*}

\begin{figure*}
    \centering
    \includegraphics[width=0.49\textwidth]{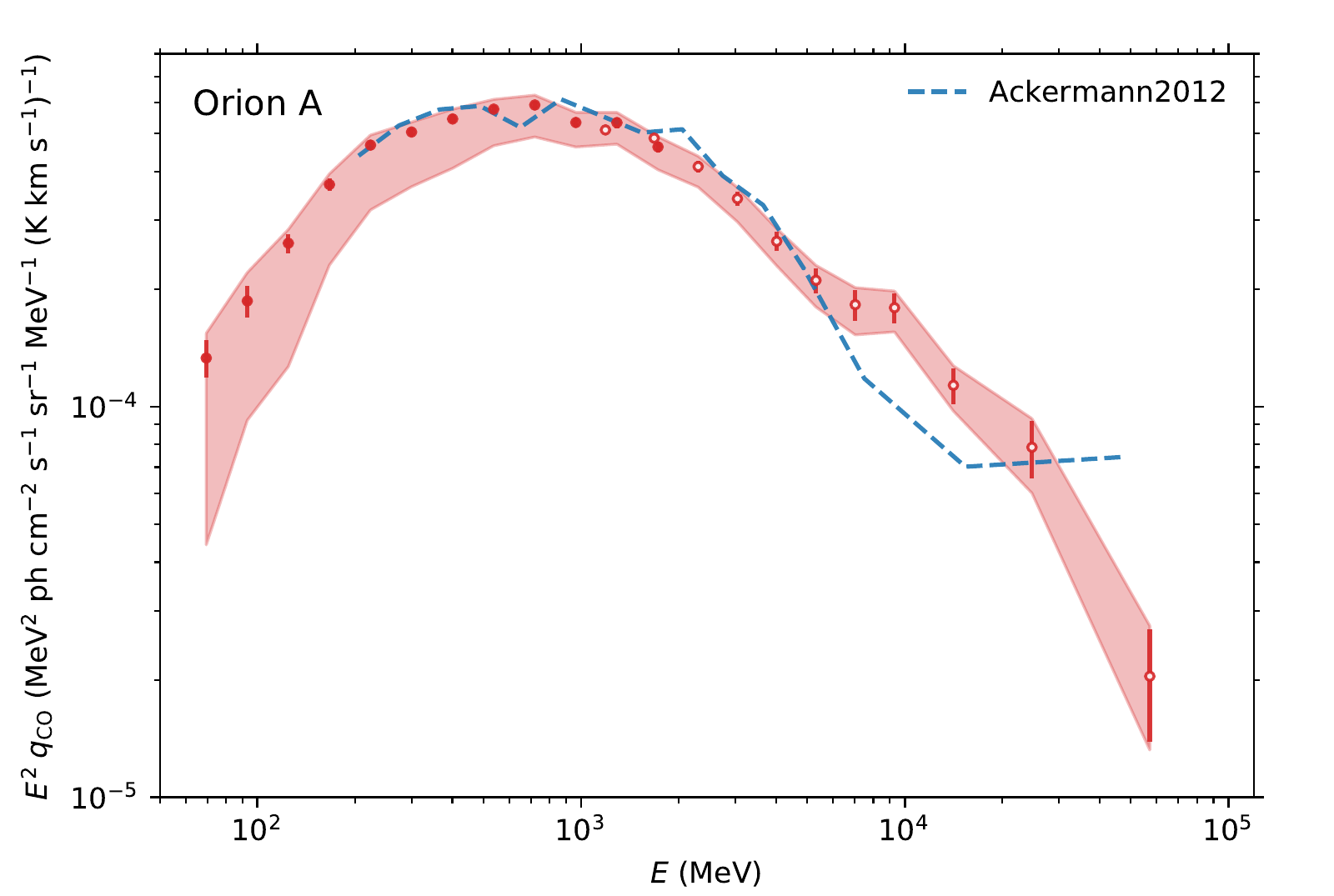}
    \includegraphics[width=0.49\textwidth]{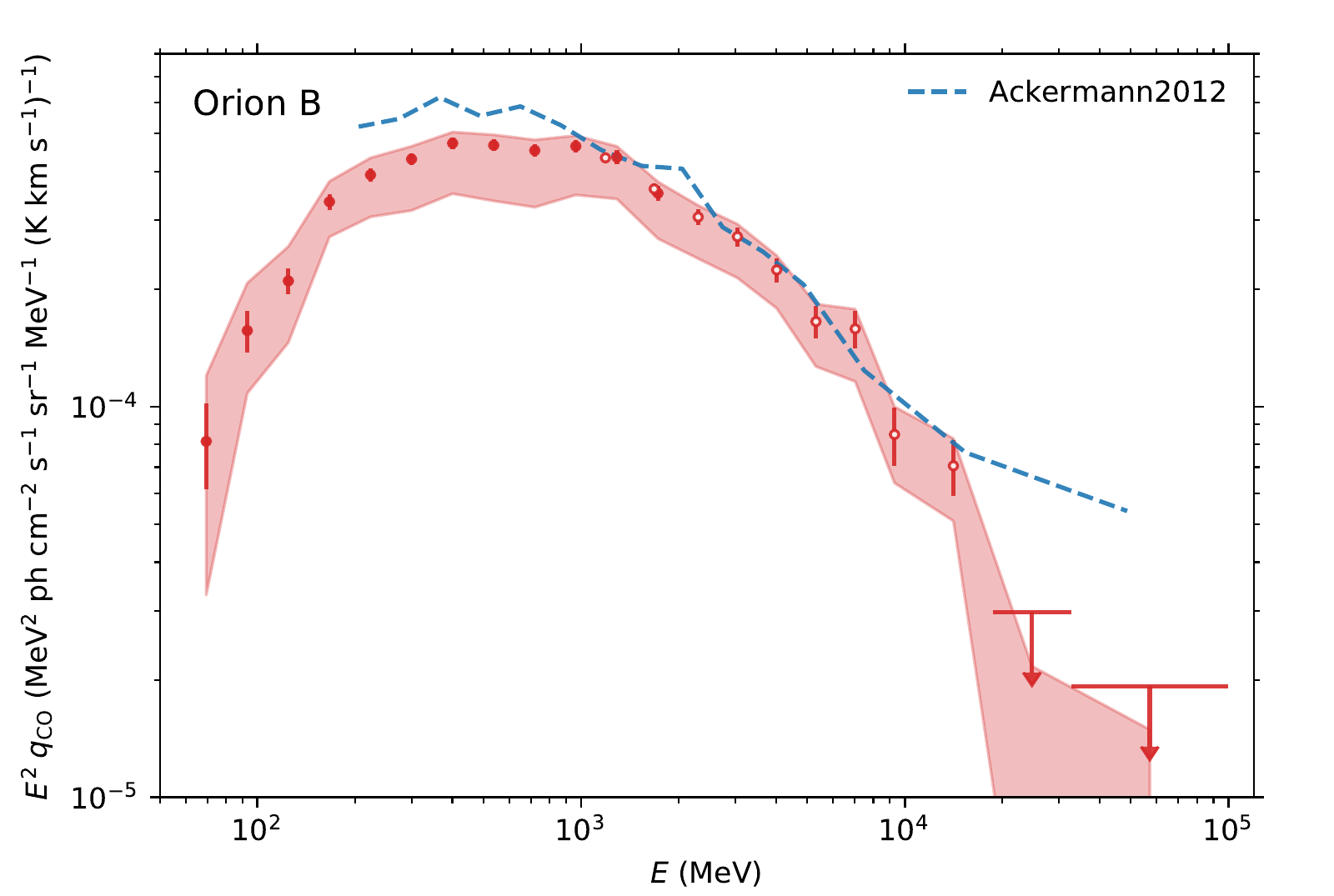}
    \caption{\label{fig:emiss}
        The emissivities of molecular clouds Orion~A and Orion~B.
        The solid and hollow circles are the emissivities derived using the baseline model for low energy and high energy data, respectively.
        The error bars correspond to the $1\sigma$ statistical uncertainties.
        The red band includes both the statistical and systematic errors.
        The 95\% upper limits are presented if the TS value in that energy bin is less than 10.
        The blue dashed lines are the emissivities for the Orion Region II (left) and Orion~B (right) components of the H$_2$-template-2 in \cite{Ackermann2012a}.
    }
\end{figure*}

We use the best-fit parameters in the bin-by-bin analysis of the baseline model to construct the expected \gr emissions in the ROI, which are further used to make residual maps for both the low energy and high energy data sets (shown in Fig.~\ref{fig:residual}).
No significant residuals except some point-like structures, which may be point sources not included in 3FGL, appear in the map.
Some weak structures remain in the residual maps, which can be caused by the inaccuracy of the adopted gas templates.

With the optimized coefficients for the molecular clouds in each energy bin, we can derive the emissivities of these sources, which are shown in Fig.~\ref{fig:emiss}.
We find the emissivities in the overlapped energy range are consistent (at most $2\sigma$ difference).
The spectral shapes of the two clouds are also consistent with each other when the uncertainties are considered, suggesting rather similar or even the same cosmic ray spectrum.
The difference in the normalizations indicates the different $X_{\rm CO}$ in the clouds.
Also plotted in the figures are the emissivities for the H$_2$-template-2 in \cite{Ackermann2012a}.
The emissivities are generally consistent with our results, and the small deviations may be caused by the different spatial templates for the target molecular clouds.
Our emissivities also match the average spectrum of all molecular clouds within $10\deg<|b|<70\deg$ reported in \cite{Casandjian2015}.

\subsection{\label{sec:analysis::systematic}Assessment of systematic uncertainties}
The emissivities of molecular clouds in Sect.~\ref{sec:analysis::gr_analysis} are calculated with the baseline templates introduced in Sect.~\ref{sec:analysis::gr_component}, and uncertainties of these models will have an impact on the results.
Besides, the systematic uncertainty on the \lat IRFs will also influence our results \citep{LatOnOrbit}.
In this subsection, we substitute the templates for the \gr radiation components and also propagate the uncertainty on effective area to understand the systematic uncertainties of our results.

The column density of \hi is calculated under the optically thin approximation in the baseline model.
However, for the regions with high \hi volume density, the opacity might not be negligible.
The 21~cm absorption spectroscopy is used to estimate the \hi opacity which is usually quantified with the spin temperature $T_{\rm S}$ \citep{HI4PI}.
According to the observations of \hi absorption, the spin temperature ranges from $\sim90$~K to $\sim400$~K \citep{Mohan2004,Ackermann2012a}, so we perform optical depth corrections by trying different $T_{\rm S}$ of 90~K, 150~K and 300~K, and derive the corresponding column density with the equation \citep{Ackermann2012b, FermiGDE}
\begin{equation}
    N_{\rm \hi}(l,b) = -C \, T_{\rm S}
        \int {\rm d}v \, \ln \left ( 1 - T_{\rm B}(l,b,v)/(T_{\rm S}-T_{\rm 0}) \right),
    \label{eqn:HI_Ts}
\end{equation}
where $T_{\rm 0}=2.66~{\rm K}$ is the brightness temperature of the background at 21~cm.
Following \cite{Ackermann2012b}, we truncate $T_{\rm B}$ to $T_{\rm S}-5~{\rm K}$ when $T_{\rm B} > T_{\rm S}-5~{\rm K}$.
The \hi column density will increase by 35.4\% (14.4\%, 6.0\%) on average in the source region for $T_{\rm S}=90~{\rm K}$ (150~K, 300~K).

Different dust tracers have different dynamical range, particularly towards the densest zones.
To test the robustness, we also adopt the \planck dust optical depth map at 353~GHz ($\tau_{353}$) \citep{Planck_tau353} as a substitute of baseline dust model.
Since a specifically tailored method is used to separate the dust emission from the cosmic infrared background, this map is significantly improved comparing with the previous $\tau_{353}$ map \citep{Planck_tau353}.
When the $E(B-V)$ cut is applied to the map, we use the conversion factor $E(B-V)/\tau_{353}=1.49 \times 10^4~{\rm mag}$ \citep{Planck_tau353_old}.

Noticing that some regions with strong dust extinction appear in the target molecular clouds, the adopt of different $E(B-V)$ clipping will change the DNM components in these regions and thus influence the spectra of the clouds.
Instead of using the threshold of $E(B-V)=5~{\rm mag}$, we make several DNM templates either without the cut or with the cut of 10~mag, 15~mag to investigate this systematic error.

The baseline IC model is calculated with the {\tt GALPROP} code for a particular propagation parameter set.
Changes of propagation parameters will lead to different spectral and spatial distributions for electrons in the Galaxy and thus different IC models.
We alter the IC model by using the templates with the GALDEF identification ${\rm ^SL ^Z4 ^R20 ^T150 ^C2}$, ${\rm ^SL ^Z10 ^R30 ^T\infty ^C5}$, ${\rm ^SS ^Z4 ^R20 ^T150 ^C2}$, and ${\rm ^SS ^Z10 ^R30 ^T\infty ^C5}$ in \cite{Ackermann2012b}.
These parameter sets replace the radial (20~kpc and 30~kpc) and vertical (4~kpc and 10~kpc) boundaries of the diffusion halo, and CR source distributions \citep{Case1998,Lorimer2006}.

The uncertainty of the effective area is the dominant part of the instrument-related systematic error, which is estimated by performing several consistency checks between the flight data and the simulated ones \citep{LatOnOrbit}.
We apply the bracketing $A_{\rm eff}$ method to propagate the systematic error to the spectra of target sources.\footnote{\url{https://fermi.gsfc.nasa.gov/ssc/data/analysis/scitools/Aeff_Systematics.html}}
Specifically, we modify $A_{\rm eff}$ to be
\begin{equation}
    A'_{\rm eff}(E,\theta) = A_{\rm eff}(E,\theta) \times \left [ 1 + \epsilon(E) \, B(E) \right ]
    \label{eqn:Aeff_sys}
\end{equation}
for all the sources except the isotropic background, the $W_{\rm CO}$ map of the Galactic Plane, the point sources with spectral indices fixed, and the template for weak point sources in the analysis.
In Eq.~(\ref{eqn:Aeff_sys}), $\epsilon(E)$ and $B(E)$ are the largest relative systematic uncertainty of the effective area and the bracketing function at energy $E$, respectively.
For both the front-converting and total data sets, if no energy dispersion correction is considered, the largest relative systematic uncertainty decreases from 15\% at 31.6~MeV to 5\% at 100~MeV, keeps 5\% between 100~MeV and 100~GeV, and then increases to 15\% at 1~TeV.
The bracketing functions $B(E) = \pm 1$ are chosen to evaluate the lower and upper bound of the spectra, while $B(E) = \pm \tanh \left [ \ln (E/E_0)/k \right ]$ with $E_0=200~{\rm MeV}$ and $k=0.32$ are used\footnote{
    No abrupt changes of $A_{\rm eff}$ with energy are expected.
    $k=0.32$ is chosen such that $B(E)$ cannot change from -0.95 to 0.95 within less than 0.5 in $\lg(E)$.
    Compared with $k=0.5$ adopted in \cite{1SC}, $k=0.32$ gives a more extreme estimate of  spectral index systematic uncertainty.
} to estimate the change of spectral index around 200~MeV.

When deriving the systematic uncertainties, we repeat the same data analysis procedure outlined in Sect.~\ref{sec:analysis::gr_analysis}.
Awaring that the uncertainties concerning the ISM are not independent since any changes in them will lead to different DNM templates and subsequently influence the target sources, we make a grid with these factors $D \in \{ A_{\rm VQ}, \tau_{353} \}$, $E(B-V)_{\rm cut}({\rm mag}) \in \{5,10,15,\infty\}$ and $T_{\rm S}({\rm K}) \in \{90, 150, 300, \infty\}$,\footnote{
    $E(B-V)_{\rm cut}=\infty~{\rm mag}$ corresponds to no $E(B-V)$ cut, and $T_{\rm S}=\infty~{\rm K}$ represents the 21~cm emission line is transparent to ISM.
} use one of the parameter combinations and evaluate the spectra.
When dealing with the IC models and $A_{\rm eff}$, we assume they are independent with other factors, so changes are only made for the parameters themselves.
Finally, we calculate the largest deviations of emissivities from the baseline one for the systematic factors related to ISM, IC and $A_{\rm eff}$, and combine them with their root sum square.\footnote{
    In the case of an upper limit, we replace the statistical error with $0.5\,(F_{\rm UL}-F_{\rm best})$ \citep{2FGL}, where $F_{\rm best}$ is the best-fit flux and $F_{\rm UL}$ is the 95\% upper limit one.
}
We find the spread of the emissivity associated with the CO gas is driven by ISM-related uncertainties below 5~GeV, and is dominated by the statistical errors in the higher energy range.

\begin{table*}
    \centering
    \caption{\label{tab:emiss}Emissivities of the molecular clouds and \hi}
    \begin{tabular*}{0.9\textwidth}{@{\extracolsep{\fill}}lccc}
  
      \hline\hline
      Energy Range\footnote{ MeV.} &
        $q_{\rm CO,OrionA}$\footnote{ \label{foot:qco}${\rm ph~cm^{-2}~s^{-1}~sr^{-1}~MeV^{-1}~(K~km~s^{-1})^{-1}}$.} &
        $q_{\rm CO,OrionB}$\textsuperscript{\ref{foot:qco}} &
        $q_{\rm \hi}$\footnote{ ${\rm ph~s^{-1}~sr^{-1}~MeV^{-1}~H^{-1}}$. The template contains both the local and distant atomic hydrogen.} \\
  
      \hline
      $   60-80    $  &  $(2.76\pm0.30^{+0.31}_{-1.82})\times10^{ -8}$  &  $(1.69\pm0.42^{+0.68}_{-0.92})\times10^{ -8}$  &  $(1.08\pm0.02^{+0.08}_{-0.48})\times10^{-28}$  \\
      $   80-107   $  &  $(2.16\pm0.20^{+0.34}_{-1.07})\times10^{ -8}$  &  $(1.81\pm0.22^{+0.54}_{-0.51})\times10^{ -8}$  &  $(8.97\pm0.19^{+0.52}_{-3.89})\times10^{-29}$  \\
      $  107-144   $  &  $(1.69\pm0.10^{+0.11}_{-0.87})\times10^{ -8}$  &  $(1.35\pm0.10^{+0.29}_{-0.40})\times10^{ -8}$  &  $(6.90\pm0.10^{+0.37}_{-3.05})\times10^{-29}$  \\
      $  144-193   $  &  $(1.33\pm0.05^{+0.07}_{-0.50})\times10^{ -8}$  &  $(1.20\pm0.06^{+0.14}_{-0.22})\times10^{ -8}$  &  $(5.07\pm0.06^{+0.27}_{-2.25})\times10^{-29}$  \\
      $  193-258   $  &  $(9.35\pm0.28^{+0.49}_{-2.96})\times10^{ -9}$  &  $(7.85\pm0.30^{+0.77}_{-1.69})\times10^{ -9}$  &  $(3.42\pm0.04^{+0.18}_{-1.49})\times10^{-29}$  \\
      $  258-346   $  &  $(5.63\pm0.16^{+0.30}_{-1.54})\times10^{ -9}$  &  $(4.80\pm0.17^{+0.34}_{-1.24})\times10^{ -9}$  &  $(2.11\pm0.03^{+0.12}_{-0.89})\times10^{-29}$  \\
      $  346-463   $  &  $(3.39\pm0.09^{+0.18}_{-0.85})\times10^{ -9}$  &  $(2.94\pm0.10^{+0.17}_{-0.75})\times10^{ -9}$  &  $(1.28\pm0.02^{+0.07}_{-0.54})\times10^{-29}$  \\
      $  463-621   $  &  $(2.00\pm0.05^{+0.11}_{-0.38})\times10^{ -9}$  &  $(1.62\pm0.06^{+0.09}_{-0.45})\times10^{ -9}$  &  $(7.33\pm0.09^{+0.43}_{-3.08})\times10^{-30}$  \\
      $  621-832   $  &  $(1.14\pm0.03^{+0.06}_{-0.19})\times10^{ -9}$  &  $(8.75\pm0.32^{+0.47}_{-2.47})\times10^{-10}$  &  $(4.14\pm0.07^{+0.25}_{-1.73})\times10^{-30}$  \\
      $  832-1114  $  &  $(5.75\pm0.18^{+0.31}_{-0.75})\times10^{-10}$  &  $(5.00\pm0.19^{+0.27}_{-1.23})\times10^{-10}$  &  $(2.16\pm0.03^{+0.14}_{-0.90})\times10^{-30}$  \\
      $ 1114-1493  $  &  $(3.20\pm0.10^{+0.17}_{-0.37})\times10^{-10}$  &  $(2.61\pm0.11^{+0.14}_{-0.56})\times10^{-10}$  &  $(1.11\pm0.02^{+0.07}_{-0.47})\times10^{-30}$  \\
      $ 1493-2000  $  &  $(1.55\pm0.05^{+0.08}_{-0.19})\times10^{-10}$  &  $(1.18\pm0.05^{+0.06}_{-0.27})\times10^{-10}$  &  $(5.50\pm0.20^{+0.34}_{-2.31})\times10^{-31}$  \\
      \hline
      $ 1000-1414  $  &  $(3.61\pm0.08^{+0.19}_{-0.51})\times10^{-10}$  &  $(3.06\pm0.09^{+0.16}_{-0.72})\times10^{-10}$  &  $(1.30\pm0.02^{+0.08}_{-0.55})\times10^{-30}$  \\
      $ 1414-2000  $  &  $(1.72\pm0.04^{+0.09}_{-0.22})\times10^{-10}$  &  $(1.28\pm0.04^{+0.07}_{-0.31})\times10^{-10}$  &  $(5.74\pm0.09^{+0.35}_{-2.41})\times10^{-31}$  \\
      $ 2000-2644  $  &  $(7.78\pm0.27^{+0.41}_{-0.85})\times10^{-11}$  &  $(5.78\pm0.27^{+0.30}_{-1.23})\times10^{-11}$  &  $(2.67\pm0.04^{+0.16}_{-1.12})\times10^{-31}$  \\
      $ 2644-3497  $  &  $(3.68\pm0.16^{+0.19}_{-0.44})\times10^{-11}$  &  $(2.94\pm0.16^{+0.15}_{-0.62})\times10^{-11}$  &  $(1.23\pm0.03^{+0.08}_{-0.53})\times10^{-31}$  \\
      $ 3497-4624  $  &  $(1.64\pm0.09^{+0.09}_{-0.20})\times10^{-11}$  &  $(1.38\pm0.10^{+0.07}_{-0.26})\times10^{-11}$  &  $(5.89\pm0.22^{+0.42}_{-2.47})\times10^{-32}$  \\
      $ 4624-6115  $  &  $(7.44\pm0.54^{+0.42}_{-0.94})\times10^{-12}$  &  $(5.84\pm0.55^{+0.31}_{-1.24})\times10^{-12}$  &  $(2.81\pm0.15^{+0.18}_{-1.18})\times10^{-32}$  \\
      $ 6115-8087  $  &  $(3.69\pm0.34^{+0.20}_{-0.50})\times10^{-12}$  &  $(3.20\pm0.36^{+0.17}_{-0.78})\times10^{-12}$  &  $(1.27\pm0.08^{+0.09}_{-0.53})\times10^{-32}$  \\
      $ 8087-10694 $  &  $(2.07\pm0.18^{+0.11}_{-0.21})\times10^{-12}$  &  $(9.81\pm1.66^{+0.52}_{-1.79})\times10^{-13}$  &  $(6.71\pm0.81^{+0.45}_{-2.99})\times10^{-33}$  \\
      $10694-18701 $  &  $(5.67\pm0.60^{+0.34}_{-0.53})\times10^{-13}$  &  $(3.53\pm0.58^{+0.21}_{-0.79})\times10^{-13}$  &  $(1.51\pm0.16^{+0.17}_{-0.65})\times10^{-33}$  \\
      $18701-32702 $  &  $(1.29\pm0.22^{+0.09}_{-0.21})\times10^{-13}$  &  $<4.88                        \times10^{-14}$  &  $(3.44\pm0.63^{+0.52}_{-1.51})\times10^{-34}$  \\
      $32702-100000$  &  $(6.25\pm2.00^{+0.83}_{-0.93})\times10^{-15}$  &  $<5.89                        \times10^{-15}$  &  $(3.44\pm0.35^{+0.35}_{-1.43})\times10^{-35}$  \\
      \hline
  
    \end{tabular*}
  
    \footnotetext{
        For the emissivities, the first uncertainties are the statistical, and the second ones are the root sum square of the systematic errors related to the ISM, IC and $A_{\rm eff}$.
        95\% upper limits of emissivities in the baseline model are given when TS$<10$.
    }
\end{table*}

The best-fit \gr emissivities of the molecular clouds including the statistical and systematic uncertainties in all energy bins are shown in Fig.~\ref{fig:emiss} and reported in Tab.~\ref{tab:emiss}.
We denote the statistical errors with vertical error bars and the total errors including systematic uncertainties with red band.

\section{\label{sec:discussion}Discussion}
Breaks at $\sim 200~{\rm MeV}$ and $\sim 2~{\rm GeV}$ present in the spectra (shown in Fig.~\ref{fig:emiss}).
To quantify the significance of the breaks, we adopt a single power law and a smoothly broken power law model to fit the data separately.
The former is defined as $q(E)=q_0\,(E/E_0)^{-\Gamma}$, while the latter is $q(E) = q_0 \, (E/E_0)^{-\Gamma_1} [1+(E/E_{\rm b})^{(\Gamma_2-\Gamma_1)/\alpha} ]^{-\alpha}$, where $\alpha$ determines the smoothness of the break and is fixed to 0.1 \citep{Ackermann2013, DAMPE2017}.
We fit the total likelihood value within a given energy range, which is the multiplication of likelihood value in each energy bin \citep{SmingTsai2013, Huang2017}.\footnote{
    The likelihood value is calculated with the interpolation of the likelihood map, which is made by scanning the likelihood value as a function of the emissivity in each energy bin.
    During the scanning, we fix the spectral parameters of point sources to the best-fit ones in the energy bin.
}
We derive the significance of the break with ${\rm TS_{br}}=-2\,\ln(\mathcal{L}_{\rm PL}/\mathcal{L}_{\rm BPL})$ according to the $\chi^2$ distribution with 2 degrees of freedom \citep{Wilks1938}, where the $\mathcal{L}_{\rm PL}$ and $\mathcal{L}_{\rm BPL}$ are the total likelihood values for the best-fit power law and smoothly broken power law models, respectively.

The data up to 1.5~GeV are used to fit the break at $\sim 200~{\rm MeV}$.
1.5~GeV is chosen as the upper bound since the high energy break happens at the energy larger than it.\footnote{
    If we change the upper bound to 1.1~GeV, the derived parameters for smoothly broken power law model are still consistent with the following ones within 2$\sigma_{\rm stat}$.
}
For Orion~A, we find a break at $E_{\rm b}=245\pm8~{\rm MeV}$ for the baseline model with a significance of $16.8\sigma$ deviating from the best-fit single power law model with a spectral index of $\Gamma=1.72\pm0.01$. 
The index of the smoothly broken power law model changes from $\Gamma_1=0.94\pm0.04$ to $\Gamma_2=1.97\pm0.03$.
For Orion~B, the optimized model shows a break at $E_{\rm b}=224\pm12~{\rm MeV}$ with the indices $\Gamma_1=0.81\pm0.07$ before the break and $\Gamma_2=1.99\pm0.06$ after.
This model is $14.0\sigma$ better than the optimized power law model with the index of $\Gamma=1.71\pm0.01$.
Taking the systematic uncertainties discussed in Sect.~\ref{sec:analysis::systematic} into account, the significance of break at the energy $245\pm8^{+9}_{-19}~{\rm MeV}$ ($224\pm12^{+13}_{-40}~{\rm MeV}$) is larger than $9.6\sigma$ ($9.0\sigma$) for Orion~A (Orion~B).\footnote{
    Different from the previous sections, the systematic errors here represent the range of the best-fit break energy in all the models analyzed in Sect.~\ref{sec:analysis::systematic}.
    The same method is applied to the systematic uncertainty of the high energy break in the following.
}

The softening at $\sim 2~{\rm GeV}$ is also evaluated with the data ranging from 464~MeV to 100~GeV.
The lower bound is chosen as 464~MeV in order to minimize the effect of low energy break in the analysis.
We have $E_{\rm b} = 1.67 \pm 0.04 ^{+0.04}_{-0.27}~{\rm GeV}$ and $1.62 \pm 0.04 ^{+0.20}_{-0.22}~{\rm GeV}$ with a significance of $10.5\sigma$ and $10.0\sigma$ for Orion~A and Orion~B, respectively.
The significances are at least $9.0\sigma$ and $8.2\sigma$ after addressing the systematic uncertainties.
The index of Orion~A (Orion~B) changes from $\Gamma_1=2.10 \pm 0.02$ ($2.07 \pm 0.02$) to $\Gamma_2=2.71 \pm 0.01$ ($2.83 \pm 0.01$) in the baseline model.

The low energy break at $\sim 200~{\rm MeV}$ is a characteristic of $\pi^0$ decay \citep{Ackermann2013}, which has also been detected in the diffuse emission of large region sky \citep{Hunter1997, Casandjian2015, FermiGDE} and some hadron-dominant supernova remnants \citep{Ackermann2013, Jogler2016, Abdalla2016}.
Our significant detection of such breaks in Orion~A and Orion~B suggests the hadronic origin of the emission.
The high energy break has also been found in other works \citep{Neronov2012, Yang2014, Casandjian2015, Neronov2017}, which has been attributed to the break of proton spectrum around $\sim10~{\rm GeV}$.

\begin{figure*}
    \centering
    \includegraphics[width=0.49\textwidth]{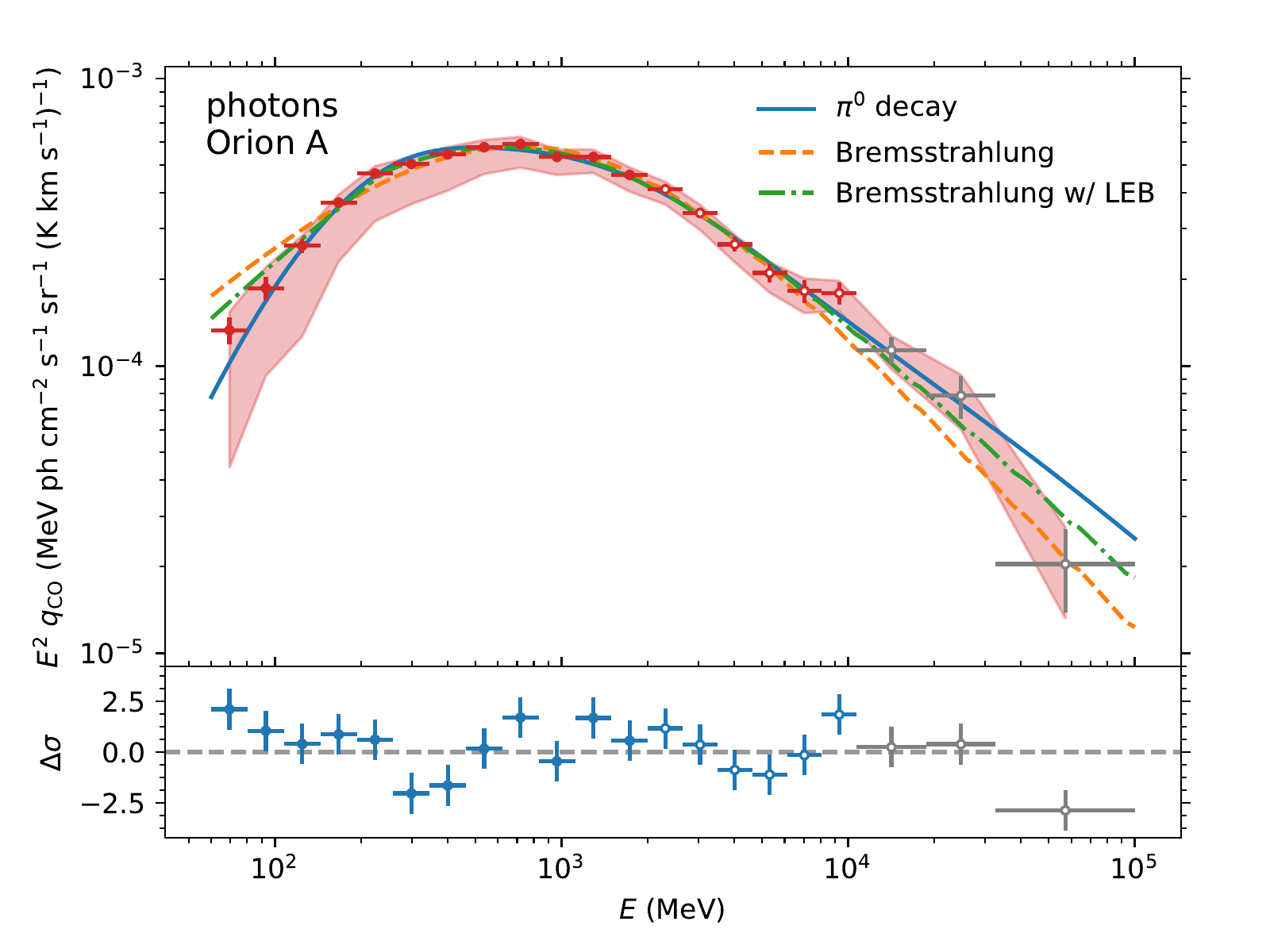}
    \includegraphics[width=0.49\textwidth]{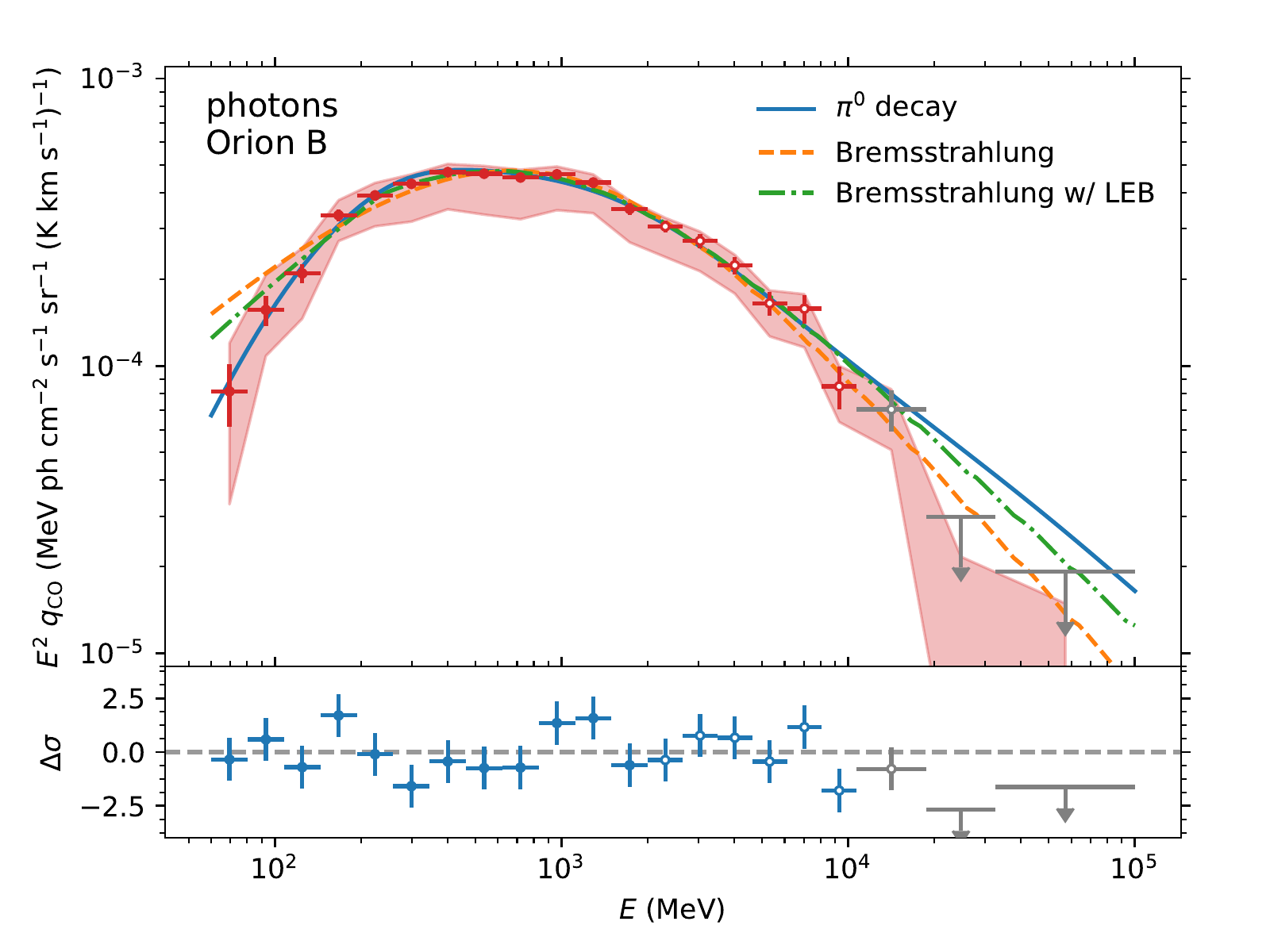}
    \caption{\label{fig:emiss_fit}
    The modeling of the \gr emissivities of Orion~A (left) and Orion~B (right).
    The data points and the error bars correspond to the emissivities and their statistical uncertainties in the baseline model, respectively.
    The red band includes both the statistical and systematic error.
    The best-fit $\pi^0$ decay and bremsstrahlung models using the data below 10.7~GeV (red points) are drawn in blue solid line and orange dashed line, respectively.
    The green dash dotted line is the bremsstrahlung model with a low energy break in electron spectrum.
    The blue points in lower panel of the figures display the residuals between the data and the best-fit $\pi^0$ decay model in standard deviation units.
    }
\end{figure*}

Below we show that the hadronic process is indeed strongly favored by data.
For such a purpose, we select the emissivities from 60~MeV to 10.7~GeV in the baseline model, fit the emissivities with the expected emissions from either the leptonic or the hadronic process.

Before calculating the emissions, we estimate the $X_{\rm CO}$ required to scale the emissivities of $W_{\rm CO}$ to the emissivities per H$_2$.
We fit the quantity $0.5\,q_{\rm CO}/q_{\rm \hi}$ with an energy independent value and simply adopt it as $X_{\rm CO}$.
The best-fit $X_{\rm CO}$ for Orion~A and Orion~B are $(1.37\pm0.01^{+0.29}_{-0.23})\times10^{20}~{\rm cm^{-2}\ (K\ km\ s^{-1})^{-1}}$ and $(1.11\pm0.01^{+0.26}_{-0.25})\times10^{20}~{\rm cm^{-2}\ (K\ km\ s^{-1})^{-1}}$ with $\chi^2=25.5$ and $27.3$ for 17 degrees of freedom, where the systematic uncertainties only include the models related to ISM.

We first fit the emissivities of the clouds with the bremsstrahlung emission.
The cross section for the electron bremsstrahlung from \cite{Strong2000} are adopted.
We also take into account the bremsstrahlung from the CR electrons and positrons scattered by the helium in the ISM, which is a factor of 0.096 the abundance of hydrogen \citep{Meyer1985}, and assume the cross section to be $\sigma_{\rm bremss,H_2}=2\,\sigma_{\rm bremss,\hi}$ \cite{Casandjian2015} .
The same electron spectral shape as \cite{Casandjian2015} is chosen, which is parameterized to be ${\rm d}F/{\rm d}E_{\rm k}=A\,(E_{\rm k}/E_0+(E_{\rm k}/E_0+P_1)^{-0.5})^{-P_2}$, where $E_{\rm k}$ is the kinetic energy and $E_0$ is 1~GeV.
After fitting the emissivities of the clouds with the model, we find the optimized spectra are softer than the observed ones below $\sim 500~{\rm MeV}$ and the models overpredict the \gr emission below 144~MeV for both clouds (see the orange dashed lines in Fig.~\ref{fig:emiss_fit}).
The $\chi^2/{\rm dof}$ for Orion~A and Orion~B are $59.1/15$ and $57.7/15$ respectively.
When a low energy break at $\sim 250~{\rm MeV}$ is introduced to the electron spectrum\footnote{
    In this case, we use a smoothly broken power law with the smoothness of the break fixed to 0.1.
}, the $\chi^2/{\rm dof}$ reduces to $24.3/13$ and $30.3/13$ for Orion~A and Orion~B, respectively (see the green dot-dashed lines in Fig.~\ref{fig:emiss_fit}).
But the data require the electron spectrum to be as hard as possible below the break, which is likely unrealistic.
Moreover, the predicted electron flux at 10~GeV is over 10 times more than the local interstellar one \citep{Cummings2016, Orlando2017}, which is also unlikely.

In the $\pi^0$ decay model, we use the cross section in \cite{Kamae2006} and the relation that $\sigma_{p{\rm H_2}}=2\,\sigma_{pp}$ \citep{Lebrun1983}.
The enhancement of 1.78 accounted for the nuclei heavier than He in both the ISM and the CR is also considered \citep{Meyer1985, Honda2004, Mori2009, Casandjian2015}.
We assume the proton spectrum at the target clouds to be ${\rm d}F/{\rm d}E_{\rm k} = A\,\beta^{P_1}\,(p/p_0)^{-P_2}$, where $p$ is the momentum of a proton \citep{Casandjian2015} and $p_0$ is fixed to $1~{\rm GeV}\,c^{-1}$.
$\beta$ is defined as $v/c$, where $v$ is the velocity.
The optimized $\pi^0$ decay model gives a reasonable fit to the both the spectra of the clouds with the $\chi^2/{\rm dof}$ to be $24.7/15$ and $19.8/15$ for Orion~A and Orion~B (see the blue solid lines in Fig.~\ref{fig:emiss_fit}).
A possible structure, however, appears in the residuals of Orion~A, which is probably contributed by the bremsstrahlung emission.
The best fit parameters of the proton spectrum are $A=(2.16^{+0.19}_{-0.15})\times10^4~{\rm m^{-2}\ s^{-1}\ sr^{-1}\ GeV^{-1}}$, $P_1=3.4^{+0.9}_{-0.7}$, $P_2=2.78^{+0.03}_{-0.03}$ for Orion~A, and $A=(2.39^{+0.24}_{-0.35})\times10^4~{\rm m^{-2}\ s^{-1}\ sr^{-1}\ GeV^{-1}}$, $P_1=3.2^{+1.0}_{-1.4}$, $P_2=2.83^{+0.03}_{-0.06}$ for Orion~B.\footnote{
    The 1$\sigma$ confidence interval $[\theta_1, \theta_2]$ is calculated by requiring $\ln \mathcal{L}(\theta_{\rm max})-0.5=\ln \mathcal{L}(\theta_i)$, where $\theta_{\rm max}$ is the best-fit parameter value \citep{Conrad2015}.
}
Our derived proton spectrum is quite similar to local interstellar one \citep{Orlando2017}, while the small difference in the normalization can be caused by the uncertainty of $X_{\rm CO}$ or enhancement factor.
Therefore, we conclude that the hadronic model can reasonably reproduce the data.

\section{\label{sec:summary}Summary}
Using approximately 107 month \lat \gr data (Fig.~\ref{fig:cmap}) and updated ISM observations including the \hi sky map from HI4PI \citep{HI4PI}, the extinction map $A_{\rm VQ}$ \citep{Planck_Avq} and optical depth map $\tau_{353}$ \citep{Planck_tau353} from \planck (see Fig.~\ref{fig:gr_gas} for the former two), we re-analyzed the molecular clouds Orion~A and Orion~B, and paid attention to the low energy emissivities.
We performed the binned likelihood analysis between 60~MeV to 100~GeV and obtained the emissivities of Orion~A and Orion~B, which are shown in Fig.~\ref{fig:emiss}.
The systematic uncertainties caused by the \gr radiation templates and the effective area are also evaluated.

Two breaks appear in the emissivities of the clouds.
In the spectrum of Orion~A, a low-energy break at $245\pm8~{\rm MeV}$ and a high-energy break at $1.67\pm0.04~{\rm GeV}$ are found at a significance level of $16.8\sigma$ and $10.5\sigma$, respectively.
For Orion~B, breaks are also found at $224\pm12~{\rm MeV}$ and $1.62\pm0.04~{\rm GeV}$ at the significance level of $14.0\sigma$ and $10.0\sigma$, respectively.
After taking into account the systematic uncertainties, the significance for the low energy breaks is still larger than $9.0\sigma$.
The break at $\sim 200~{\rm MeV}$, recognized as a signature of the $\pi^0$ decay emission \citep{Ackermann2013}, has been revealed {\it for the first time}, providing a direct evidence for the hadronic origin of the \gr emission from the molecular clouds.

Finally, we would like to point out that the low energy \gr emissions of Orion~A and Orion~B are different from the so-called Galactic GeV excess, which may challenge the molecular cloud origin model \citep{deBoer2017} for the latter.

\begin{acknowledgments}
    We appreciate R.~Yang, Z.-H.~He, Y.-J.~Li, N.-H.~Liao, Y.-H.~Ma, X.-L.~Wang and Z.-Q.~Xia for the helpful discussions.
    In this work, we have also used {\tt NumPy} \citep{numpy}, {\tt SciPy}\footnote{\url{http://www.scipy.org}}, {\tt Matplotlib} \citep{matplotlib}, {\tt Astropy} \citep{astropy} and {\tt iminuit}\footnote{\url{https://github.com/iminuit/iminuit}}.
    This work is supported by National Key Program for Research and Development (2016YFA0400200), the National Natural Science Foundation of China (Nos. 11433009, 11525313, 11722328), and the 100 Talents program of Chinese Academy of Sciences.
\end{acknowledgments}

\end{document}